\documentclass{iauc}[12pt]

\usepackage{graphicx}
\pubyear{2005}
\volume{199}
\pagerange{1--}
\jname{Probing Galaxies through Quasar Absorption Lines}
\editors{P. R. Williams, C. Shu, and B. M\'{e}nard, eds.}
\title[Spectroscopic Studies of \boldmath{$z \sim5.7$} and $z\sim6.5$ Galaxy Samples] 
{Spectroscopic Studies of \boldmath{$z \sim5.7$} and $z\sim6.5$ Galaxies: Implications for Reionization}

\author[Hu et al.]   
{Esther M. Hu, Lennox L. Cowie, Peter Capak%
  \thanks{Present address: Department of Astronomy, California Institute of Technology, MS 105-24, 1201 E. California Blvd., Pasadena, CA, 91125, USA.},
\and Yuko Kakazu}

\affiliation{Institute for Astronomy, University of Hawaii,
  2680 Woodlawn Drive, Honolulu, HI 96822, USA}

\begin{document}

\maketitle

\begin{abstract}
The recent development of large, complete samples which identify
high-redshift galaxies at $z\sim5.7$ and $z\sim 6.5$ from deep,
wide-field surveys provides detailed information on the earliest
galaxies, their numbers, spatial and kinematic distributions,
and implications for early reionization of the IGM.   In this
contribution we present results of spectroscopic studies of $z\sim
5.7$ and $z\sim6.5$ galaxies identified from our deep, Lyman alpha
narrowband and multicolor surveys conducted with the SuprimeCam
mosaic CCD camera on the 8.3-m Subaru telescope and observed
with the DEIMOS multi-object spectrograph on Keck. The luminosity
function of the $z\sim6.5$ galaxies is shown to  be similar to the
luminosity function of the $z\sim 5.7$ galaxy samples, suggesting
that a substantial star-forming population is already in place at
$z\sim 6.5$.  Comparisons of both individual and stacked spectra of
galaxies in these two samples show that the Lyman alpha emission
profiles, equivalent widths, and continuum break strengths do not
substantially change over this redshift interval.  The wide-field
nature of the surveys also permits mapping the large-scale
distribution of the high-redshift galaxies in spatial structures
extending across individual SuprimeCam fields ($\sim 60$ Mpc).
Field-to-field variations in the number of objects at $z\sim 6.5$
may shortly be able to place constraints on the porosity of the
reionization boundary.

\keywords{cosmology: observations --- early universe --- galaxies: distances
          and redshifts --- galaxies: evolution --- galaxies: formation --- galaxies: high-redshift}


\end{abstract}
\firstsection 
\section{Introduction}
Substantial progress has been made over the last 5 years in
high-redshift galaxy studies (see \cite{spinrad04} for a review).
The significant aspects of recent work carried out since that review
are the rapid growth in the numbers of galaxies that populate
the high-redshift end of the samples ($z\sim6$ and above)  and
the development of large, spectroscopically complete samples from
wide-area surveys for a number of discrete fields (e.g., \cite{lum5},
\cite{tani}, \cite{hdfz5}).  These  give a handle on the problem
of cosmic variance, and also allow us to build up the large samples
required for luminosity function studies. Spectroscopic completeness
makes it  possible to estimate the star-forming population at high
redshift, and its evolution.  Detailed examination and comparison
of the strength and profile shape of the Ly$\alpha$ emission line
may provide insights to possible changes in the underlying stellar
population, or in the neutral hydrogen content of the surrounding
intragalactic medium.

\section{The Hawaii Wide-Field Narrowband Surveys}
The most successful method of identifying galaxies at redshifts
beyond $z\sim 5.5$, where galaxy continuua are faint against a
strong nightsky background, is by using the redshifted Ly$\alpha$
emission line.  The present surveys were were driven by the
development of wide-field, red-sensitive instruments on the large
telescopes: the half-degree FOV SuprimeCam mosaic CCD imager
(Miyazaki et al.\ 2002) on the 8.3-m Subaru Telescope, and the
wide-field ($16.7'\times 5'$) DEIMOS multi-slit spectrograph
and imager on the Keck II 10-m telescope.  Narrowband imaging is
used in combination with deep multicolor $BVRIz'$ and moderate
infrared $K'$ imaging over the SuprimeCam fields to identify
high-redshift galaxies with strong Ly$\alpha$ emission and
continuum color breaks. 

Current investigations use $\sim 120$\AA\
bandpass filters centered at 8150\AA\ and 9130\AA\ (Ly$\alpha$
at  $z\sim 5.7$  and $z\sim 6.5$) for the narrowband studies, with
follow-up DEIMOS spectra (3.6\AA\ resolution, $R\sim2700$) at Keck.
Survey fields are chosen from half a dozen extensively studied
fields: SSA22, GOODS-N, Lockman Hole NW, SSA13, SSA17, and A370.
Completed studies at $z\sim 5.7$ for SSA22 (\cite{lum5}) show a
high success rate in identifying and confirming candidates (19
confirmed of 22 observed), as well as  success in discriminating
against red stars and identifying/confirming T dwarf candidates
(\cite{kakazu}). A second $z\sim5.7$ study has been completed for
the region including the HDF-N and GOODS-N fields (\cite{hdfz5}).
At present 62 confirmed redshifts at $z\sim5.7$ are in hand for
3 fields, with 14 confirmed $z\sim6.5$ redshifts (extending to
$z=6.74$) for 3 fields.

\begin{figure}[tp]
\includegraphics[width=5.2in]{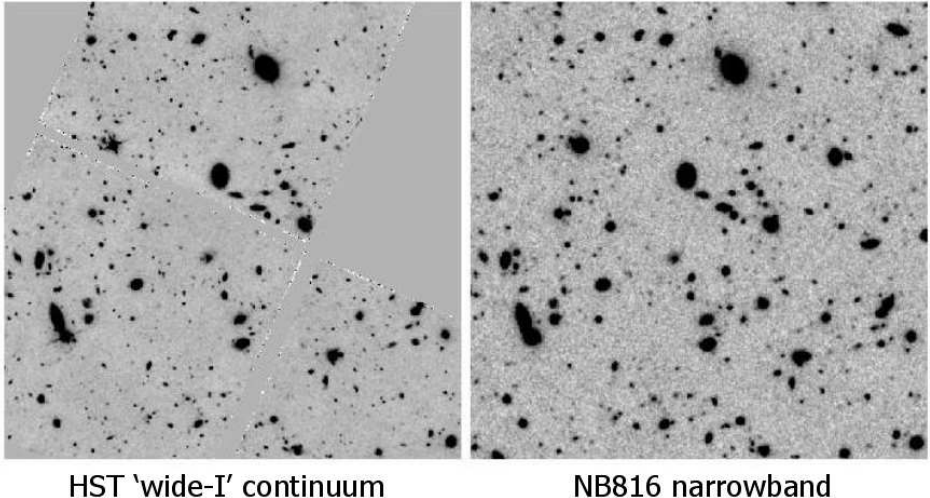}
\caption{({\it Left panel\/}) shows the Hubble Deep Field (HDF-N)
imaged with the WFPC2 camera on HST through the F814W continuum
filter.  ({\it Right panel\/}) shows the same region imaged from
the ground through a $120\,$\AA-wide narrowband NB816 filter (for
Ly$\alpha$ emission at $z\sim 5.7$) using the SuprimeCam mosaic
CCD camera on Subaru.  The narrowband images in the survey are
comparable in depth to the HDF, and each field center surveys a
region 70 times larger than the area shown here, with good image
quality. Wide area coverage and image depth are required to identify
and study the high-redshift samples.}\label{fig:hdf_nb_comp}
\end{figure}
At magnitudes brighter than $AB\sim25.5$ the number density of such
high-redshift galaxies is only a few hundred per square degree and
the distribution is highly correlated on sub-degree scales (e.g.,
Capak 2004). A combination of depth and area coverage is needed
for these studies. Figure~\ref{fig:hdf_nb_comp} shows the depth 
and image quality of the survey's SuprimeCam narrowband exposures
compared to deep HDF continuum images. No $z\sim5.7$ galaxies
are found in the survey within the HDF-N.  Over the larger
ACS GOODS-N field (still a sub-region of the SuprimeCam
field) only 6 $z\sim5.7$ galaxies were found. And none of the
$z\sim6.5$ sources were found within the region of ACS GOODS-N.

\section{Luminosity Functions}
\begin{figure}
\centering
\includegraphics[width=4.2in,scale =0.9]{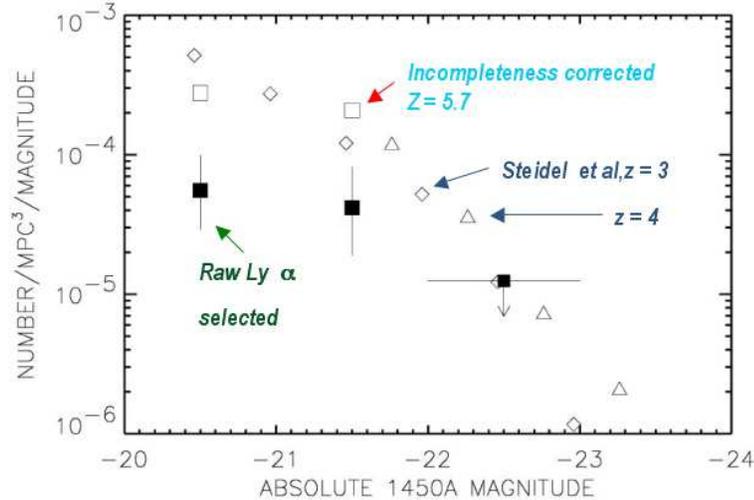}
\caption{The UV continuum luminosity function of  identified
$z\sim 5.7$ galaxies ({\it boxes\/}) compared with the UV continuum
luminosity functions at $z\sim3$ ({\it diamonds}) and at $z\sim 4$
reported by \cite{stei1999}.  Measured points are shown as {\it
solid boxes} with $1\sigma$ Poisson uncertainties based on the
number of objects in each bin; the open boxes that match closely
to the lower redshift luminosity functions show assumed values if
emitters pick out 20\% of the LBG sample as \cite{stei2000} find
to be the case at $z\sim3$.}\label{fig:uvlum}
\end{figure}

With the high-redshift samples in hand, we can construct luminosity
functions at $z\sim5.7$ and $z\sim6.5$, and compare these to
results at lower redshifts.  The most direct comparison is to use
the UV continuum luminosities of the Ly$\alpha$-selected samples and
compare these with luminosity functions of the Lyman-break galaxies
(\cite{stei1999}, \cite{stei2000}) at redshifts $z\sim3$ and $z\sim4$
(Figure~\ref{fig:uvlum}).  This avoids problems of estimating
possible dust contamination on the emission line.
Here the main uncertainty is how to
correct the raw counts for the fractional population with strong
Ly$\alpha$ emission.  If the estimate of \cite{stei2000}
based on the LBGs at $z\sim3-4$ with strong Ly$\alpha$ is used
to assume that the Ly$\alpha$ emitters represent only 20\% of
the star-forming galaxy population at $z\sim5.7$, then the raw
numbers for Ly$\alpha$-selected galaxies ({\it filled boxes\/})
are corrected upwards to the {\it open boxes\/}  that overlie the
$z\sim3$ and $z\sim4$ Lyman break galaxy UV continuum luminosity
functions, implying there has not been a substantial decrease in
the star formation rate  out to redshift $z\sim5.7$. 

\begin{figure}
\centering
\includegraphics[width=4.2in,scale =0.9]{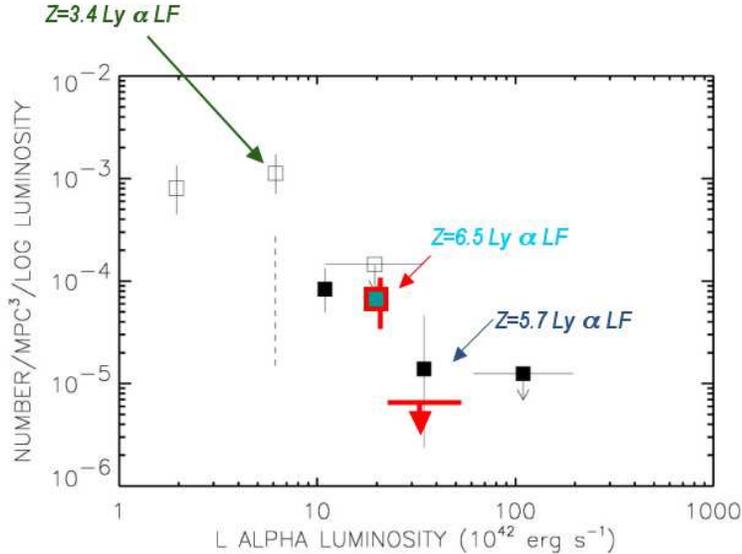}
\caption{The Ly$\alpha$ luminosity function of  identified
$z\sim 5.7$ galaxies ({\it filled boxes\/}) compared with the Ly$\alpha$
luminosity functions at $z\sim3.4$ ({\it open boxes}) 
(\cite{cow98}, \cite{smitty}).  Measured points are shown with $1\sigma$ Poisson uncertainties based on the
number of objects in each bin.  The vertical dashed line indicates the limits of
the $z\sim5.7$ survey, which does not extend to the faint end of the $z\sim3.4$
Ly$\alpha$ luminosity function. The points and upper limits for the $z\sim6.5$
luminosity function are shown with the thick bars, and are consistent with the
$z\sim5.7$ luminosity function, but are based on only 14 galaxies.}\label{fig:lalum}
\end{figure}

For the $z\sim6.5$ sample the longest wavelength deep continuum band
($z'$ around  9200 \AA) spans the wavelength of Ly$\alpha$ emission, making
it difficult to estimate the line-free ultraviolet continuum.
Here, we compare the Ly$\alpha$ luminosity function with
the luminosity function of the emission line obtained in narrowband samples
at $z\sim3.4$ (\cite{cow98}, \cite{smitty}).
The results are shown in Figure~\ref{fig:lalum}.  The $z\sim 3.4$ points
are plotted with {\it open boxes\/} and the $z\sim 5.7$ points are shown
with {\it filled boxes\/}.  The vertical dashed line indicates the limits of
the sample, which does not reach the flat part of the $z\sim5.7$
luminosity function. The {\it filled box and upper limit drawn with heavy bars\/}
represent the points for the $z\sim6.5$ luminosity function and overlie the
$z\sim5.7$ luminosity function data.  The points
are consistent with no marked change in the luminosity function
between $z\sim5.7$ and $z\sim6.5$ as suggested by \cite{mal}, who used
a likelihood analysis to merge disparate data samples with varying
degrees of incompleteness to compare the two redshift samples.
The advantage of the present results lies in the larger and
homogeneous samples, and the completeness of identifications.

\section{Ly$\alpha$ Emission Line Profiles}
\begin{figure}\hskip1.6in
\centering
\includegraphics[width=3.3in,angle=90,scale =0.9]{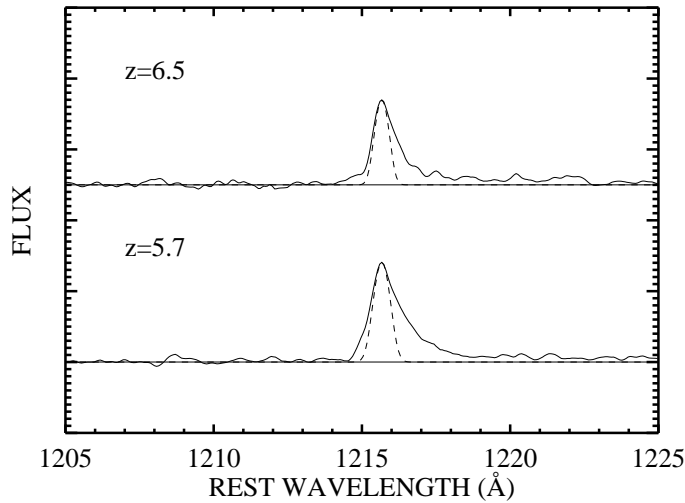}
\vspace{-0.25in}
\caption{ Stacked profiles of identified $z\sim 6.5$ galaxies
and $z\sim5.7$ galaxies.  The dotted profile shows the spectral
resolution from the night sky lines, and the steep blue fall-off
due to absorption by neutral hydrogen is clearly evident in the
asymmetric line profile. The profiles are extremely similar in
equivalent width (56\AA\ and 60\AA) and shape.  This may suggest
that the reionization epoch is at higher redshift, since we might
hope to see a change in the profiles due to the expected increase
in scattering redward of Ly$\alpha$ from hydrogen damping wings
due to the surrounding neutral medium if we sample galaxies above
and below the reionization redshift (suggested at $z\sim 6.2$).}\label{fig:z5z6comp}
\end{figure}

If reionization occurred at redshifts lower than
$z=6.5$ we should see a major change in the equivalent widths
of the Lyman alpha emission line between the $z=6.5$ sample and
the $z=5.7$ sample since the damping wings of the neutral gas
should scatter much of the redward wing of the Lyman alpha
lines at the higher redshifts where the intergalactic gas
is neutral. We can examine the emission-line
profiles in the two samples. Figure~\ref{fig:z5z6comp} shows that
the Ly$\alpha$ profiles, which have been stacked to improve
our sensitivity, are extremely similar. Both profiles show the
strong blue asymmetry due to the Ly$\alpha$ forest that is
characteristic of the very high-redshift galaxies. The profiles have
similar equivalent widths and shapes.

Does this provide clear evidence that the epoch of reionization must lie
at higher redshifts?  Not necessarily.  For Ly$\alpha$-emitting galaxies
that are sufficiently luminous, self-clearing can complicate their effectiveness
as a probe of the surrounding neutral IGM (e.g., \cite{haiman}).  
Even for lower luminosity objects, clustered distributions or neighboring
objects may ionize the local region. The $z\sim 6.5$ samples are
still relatively small, and show a surprising amount of field-to-field
variation, with 10 of the identified systems coming from the A370 field --
coincidentally, the first field with a discovered $z>6$ galaxy (\cite{z6}).
Preferred (low density) lines of sight could be ionized, and there
can be selection biases in our $z\sim 6.5$ sample.
The properties could reflect large-scale structure effects
or a porous reionization boundary in both the profile and
luminosity function analyses.

\section{Large-Scale Structure}
\begin{figure}[ht]
\includegraphics[angle=90, width=2.3in]{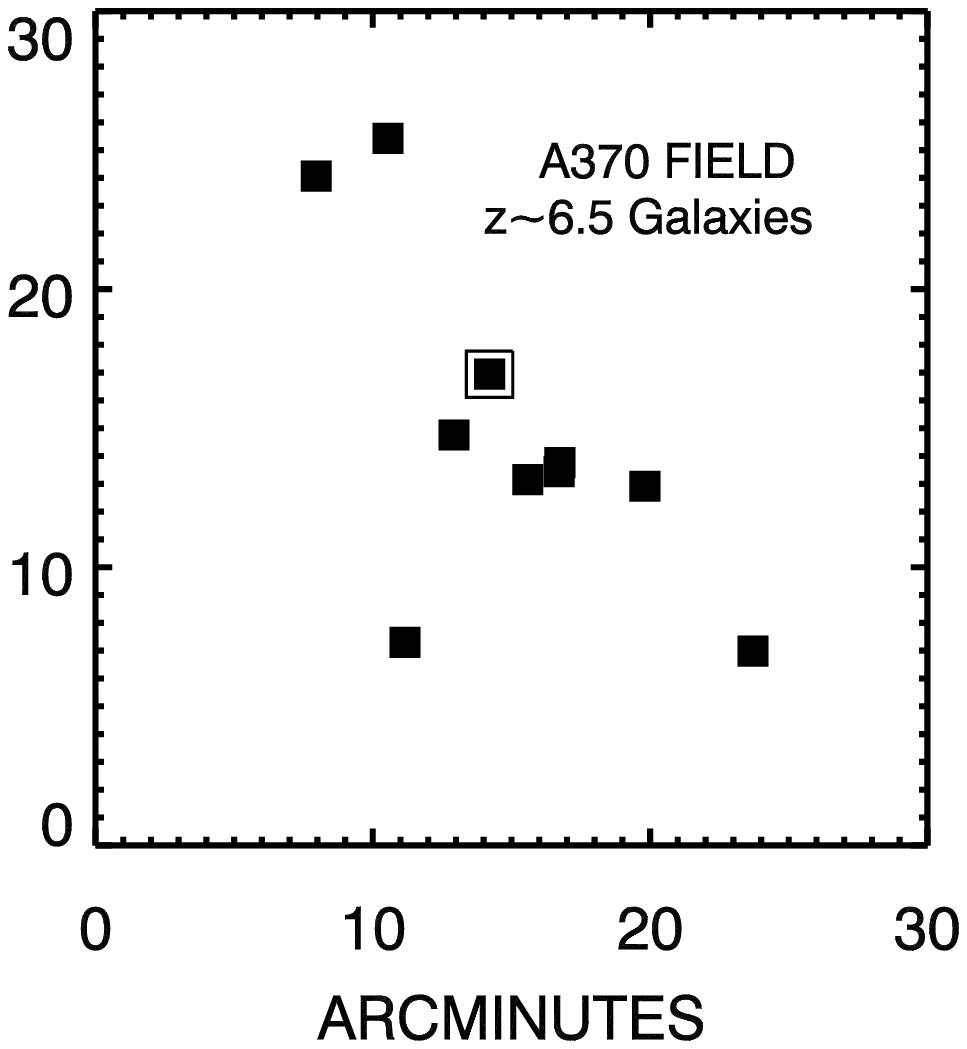}\hskip 0.3in
\includegraphics[width=2.3in]{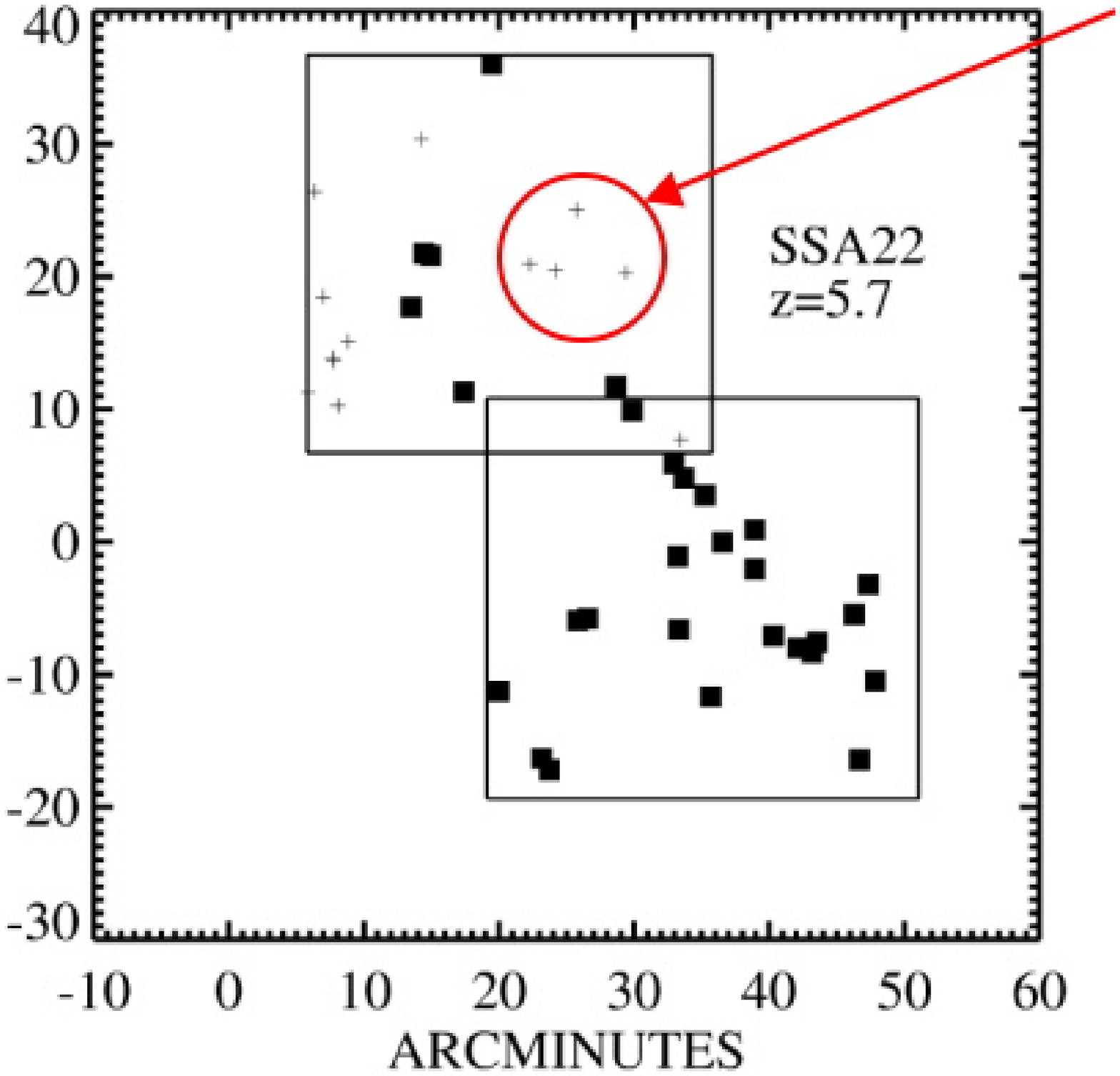}
\caption{({\it Left})The $z\sim6.55$ emitters in A370 display a filamentary
   extent.  The box
   indicates a $z\sim 6.74$ galaxy, currently, the highest confirmed
   redshift. ({\it Right}) Redshift $z\sim5.7$ structure in the expanded SSA22 field.
   Small crosses (circled region) show candidates not yet spectroscopically observed
   along the extended filament.
  }\label{fig:structure}
\end{figure}

The $z\sim5.7$ spectroscopic samples (\cite{lum5}, \cite{hdfz5}) with near-complete coverage in the SSA22 and HDF-N SuprimeCam
fields provided the first quantitative demonstration of  the existence of structured kinematic and redshift  structures on large scales ($\sim 60$ Mpc co-moving
distance) at these redshifts. The long diagonal filament in {\it right panel\/}
of Figure~\ref{fig:structure} makes up a distinct redshift system with Ly$\alpha$
emission falling in the bottom quarter of the filter.  Such features are also seen in
maps of the HDF-N. Similar structured distribution
at $z\sim 5.7$ has recently been reported by \cite{ouchi} for the Subaru XMM Deep
Field.  

\section{Conclusions}
We are now able to obtain large samples of $z\sim5.7$ and $z\sim6.5$  galaxies.
The continuum and Ly$\alpha$ luminosity functions seem to be similar to those
at lower redshifts, indicating a strong contribution to to the ionization of the 
IGM from star-formation at these redshifts. Both the Ly$\alpha$ luminosity function and
the Ly$alpha$ line shapes are similar at $z\sim5.7$ and $z\sim6.5$. Large scale
structure revealed in the distribution of these high-redshift galaxies leads us to the
exciting conclusion that we may
be able to make 3-dimensional maps of the cosmic web at these redshifts!

\begin{acknowledgments}
This research was supported by NSF grants AST 00-71208 to E.\,M.\,H.
and AST99-84816 to L.\,L.\,C., and by NASA grant GO-7266.01-96A from Space Telescope Science Institute, which is operated by AURA, Inc. under NASA contract NAS 5-26555. We are grateful to the staff of the
Keck and Subaru Telescopes for supporting these observations.
\end{acknowledgments}

\end{document}